\documentclass[prb,onecolumn,showpacs,preprintnumbers,amsmath,amssymb]{revtex4}

\usepackage{epsfig}
\usepackage{dcolumn}
\usepackage{bm}
\newcommand{\mmty}{Hg$_{\scriptstyle 1-y}$Mn$_{\scriptstyle y}$Te\,}
\newcommand{\mmt}{Hg$_{\scriptstyle 0.98}$Mn$_{\scriptstyle 0.02}$Te\,}
\newcommand{\cmtx}{Hg$_{\scriptstyle 1-x}$Cd$_{\scriptstyle x}$Te\,}
\newcommand{\cmt}{Hg$_{\scriptstyle 0.3}$Cd$_{\scriptstyle 0.7}$Te\,}

\DeclareMathAlphabet{\mathitb}{OT1}{cmr}{bx}{sl}

\begin{document}


\title{Band structure of semimagnetic Hg$_{1-y}$Mn$_y$Te quantum wells}

\author{E. G. Novik$^1$}
\email{novik@physik.uni-wuerzburg.de}
\author{A. Pfeuffer-Jeschke$^1$}
\author{T. Jungwirth$^{2,}$\footnote{Also at School of Physics and Astronomy, University of
Nottingham, Nottingham NG7 2RD, UK; and Physics Department,
University of Texas at Austin, Austin, Texas 78712-0264, USA.}}
\author{V. Latussek$^1$}
\author{C. R. Becker$^1$}
\author{G. Landwehr$^1$}
\author{H. Buhmann$^1$}
\author{L. W. Molenkamp$^1$}
\affiliation{$^1$Physikalisches Institut der Universit\"{a}t
W\"{u}rzburg, Am Hubland, 97074 W\"{u}rzburg, Germany \\
$^2$Institute of Physics ASCR, Cukrovarnick\'{a} 10, 162 53 Praha
6, Czech Republic}

\date{\today}

\begin{abstract}
The band structure of semimagnetic
Hg$_{1-y}$Mn$_y$Te/Hg$_{1-x}$Cd$_x$Te type-III quantum wells has
been calculated using eight-band ${\mathitb k}\cdot {\mathitb p}$
model in an envelope function approach. Details of the band
structure calculations are given for the Mn free case ($y=0$). A
mean field approach is used to take the influence of the $sp-d$
exchange interaction on the band structure of QW's with low Mn
concentrations into account. The calculated Landau level fan
diagram and the density of states of a
Hg$_{0.98}$Mn$_{0.02}$Te/Hg$_{0.3}$Cd$_{0.7}$Te QW are in good
agreement with recent experimental transport observations. The
model can be used to interpret the mutual influence of the
two-dimensional confinement and the $sp-d$ exchange interaction on
the transport properties of Hg$_{1-y}$Mn$_y$Te/Hg$_{1-x}$Cd$_x$Te
QW's.

\end{abstract}



\maketitle

\section{\label{sec:level1}Introduction}

Recently, numerous spin related observations have been published
which involve optical and transport experiments on diluted
magnetic, semiconducting heterostructures and quantum wells
(QW's).\cite{Hayashi98,Scherbakov01,Gui04,Camilleri01,Awschalom99}
The correct interpretation of these effects requires a detailed
knowledge of the underlying band structure. This is especially
important for narrow gap semiconductors,\cite{PhD} because strong
band mixing prevents a simple interpretation of optical and
transport results by means of a parabolic band model which might
be still applicable for most wide gap materials such as GaAs or
InGaAs.\cite{Bastard88,Das90}

Here, we concentrate on the band structure calculations of
HgTe/\cmtx{} quantum wells. This material has some interesting
properties: depending on the QW width ($d_W$), the QW has either a
{\it normal} or {\it inverted} band structure when $d_W<6$~nm or
$d_W>6$~nm, respectively. In the latter case the conduction band
exhibits $\Gamma_8$ symmetry which leads to a strong Rashba
spin-orbit splitting in QW's with an asymmetrical confinement
potential.\cite{PhD,Zhang01} Additionally, the spin splitting of
the subbands can be enhanced by introducing magnetic ions (Mn) in
the QW structure, e.g., \mmty{}/\cmtx{}. It should be noted that
in II-VI semiconductors, Mn is incorporated into the crystal
lattice isoelectrically and does not act as a donor or an
acceptor. Therefore, Mn ions act primarily as a magnetic but not
as a Coulomb impurity and mobilities achieved for these QW
structures  with low Mn concentrations are comparable with those
for non-magnetic structures.

This paper is organized as follows: In section II a detailed
description of the model used for the band structure calculations
is presented. In subsection II-A, we consider the model for
non-magnetic as well as for magnetic QW's at zero external
magnetic field. This model is used to calculate the subband energy
dispersion of \mmt{}/\cmt{} and HgTe/\cmt{} QW's. In subsection
II-B, the band structure model for a HgTe/\cmtx{} QW in an
external magnetic field is described. This model is extended in
subsection II-C in order to take the influence of the $sp-d$
exchange interaction on the band structure of magnetic QW's into
account. The Landau level fan diagram and the density of states of
\mmt{}/\cmt{} QW are compared with recent transport experiments,
and section III summarizes the results. In the appendices details
of the calculations for different growth directions (a) and strain
as well as piezoelectric effects (b) are discussed.

\section{\label{sec:level1} Band structure model}

\subsection{\label{sec:level2} $B=0$}

The band structure model we use is based on an envelope function
approximation introduced by Burt.\cite{Burt99} The total wave
function is expanded in terms of band edge ($\mathitb{k}=0$) Bloch
functions $u_{n}$:

\begin{equation}\label{Psi}
  \Psi(\mathitb{r})=\sum_{n}F_{n}(\mathitb{r})~
  u_{n}(\mathitb{r}),
\end{equation}
where $F_{n}(\mathitb{r})$ are the envelope functions. $u_{n}$ is
assumed to be the same in the barrier and the well layers.
Assuming translation invariance in the plane perpendicular to the
growth direction ($z$ axis) $F_{n}$ can be represented as follows
\begin{equation}\label{Envelope}
  F_{n}=\exp[i(k_{x}x+k_{y}y)]~f_{n}(z),
\end{equation}
where $k_{x}$ and $k_{y}$ are the wave vector components in the
plane of the QW. The envelope functions and the energy levels near
$\mathitb{k} = 0$ are determined within the framework of
$\mathitb{k}\cdot\mathitb{p}$ theory by solving a system of
coupled differential equations:\cite{Kane66,Kane57}
\begin{eqnarray}\label{Eignprob}
  & & \sum_{n'}\left(H_{nn'}+V(z)\delta_{nn'}\right)f_{n'}(z)= \nonumber\\
  & & \sum_{n'}\left(\sum_{\alpha,\beta}^{x,y,z}
k_{\alpha}D_{nn'}^{\alpha\beta}k_{\beta}
 +\sum_{\alpha}^{x,y,z}P_{nn'}^{\alpha}k_{\alpha}+E_{n'}(z)\delta_{nn'} +
V(z)\delta_{nn'}\right)f_{n'}(z)=E\cdot f_{n}(z),
\end{eqnarray}
where  $n$ and $n'$ are the summation indices for the sum over the
dimensionality of the chosen basis set, $E_{n'}(z)$ are the
respective band edge potentials, and $V(z)$ is the
self-consistently calculated Hartree-potential. The momentum
matrix elements $P_{nn'}^{\alpha}$, which describe the coupling
between the $n$ and $n'$ bands, are treated exactly, while the
$D_{nn'}^{\alpha\beta}$, which describe the perturbative coupling
of the $n$ and $n'$ bands to the remote bands, are calculated
using second-order perturbation theory.
\cite{Luttinger56,Pidgeon66}

In narrow-gap HgTe based structures, the strong coupling between
$s$-like conduction and $p$-like valence bands causes mixing of
the electronic states and induces nonparabolicity in the
conduction bands. These effects were taken into account exactly by
Kane\cite{Kane57} in the framework of the
$\mathitb{k}\cdot\mathitb{p}$ theory. In order to consider the
coupling between the $\Gamma_6$, $\Gamma_7$ and $\Gamma_8$ bands
we choose the usual eight-band basis set (see
Refs.~\onlinecite{Kane66} and \onlinecite{Winkler03}):
\begin{eqnarray}\label{BasisSet}
  u_{1}(\mathitb{r})=|\Gamma_6,+1/2\rangle &=&  S\uparrow \nonumber\\
  u_{2}(\mathitb{r})=|\Gamma_6,-1/2\rangle &=&  S\downarrow \nonumber\\
  u_{3}(\mathitb{r})=|\Gamma_8,+3/2\rangle &=&
1/\sqrt{2} (X+iY)\uparrow \nonumber\\
  u_{4}(\mathitb{r})=|\Gamma_8,+1/2\rangle &=&
1/\sqrt{6} [(X+iY)\downarrow -2Z\uparrow] \nonumber\\
  u_{5}(\mathitb{r})=|\Gamma_8,-1/2\rangle &=&
-1/\sqrt{6} [(X-iY)\uparrow   +2Z\downarrow] \\
  u_{6}(\mathitb{r})=|\Gamma_8,-3/2\rangle &=&
-1/\sqrt{2} (X-iY)\downarrow \nonumber\\
  u_{7}(\mathitb{r})=|\Gamma_7,+1/2\rangle &=&
1/\sqrt{3} [(X+iY)\downarrow +Z\uparrow] \nonumber\\
  u_{8}(\mathitb{r})=|\Gamma_7,-1/2\rangle &=&
1/\sqrt{3} [(X-iY)\uparrow -Z\downarrow]
  \nonumber.
\end{eqnarray}
The total angular momentum is then given by $j=1/2$ or $j=3/2$.

For the chosen basis set, the Hamiltonian $H_{nn'}$ in
Eq.~\ref{Eignprob} for a two-dimensional system with [001] growth
direction takes the following form:\cite{PhD}
\begin{equation}\label{Hamilt}
H =
\begin{pmatrix}
  T & 0 & -\frac{1}{\sqrt{2}}Pk_{+} & \sqrt{\frac{2}{3}}Pk_{z} &
\frac{1}{\sqrt{6}}Pk_{-} & 0 & -\frac{1}{\sqrt{3}}Pk_{z} &
-\frac{1}{\sqrt{3}}Pk_{-} \\
  0 & T & 0 & -\frac{1}{\sqrt{6}}Pk_{+} & \sqrt{\frac{2}{3}}Pk_{z} &
\frac{1}{\sqrt{2}}Pk_{-}  & -\frac{1}{\sqrt{3}}Pk_{+} &
\frac{1}{\sqrt{3}}Pk_{z} \\
  -\frac{1}{\sqrt{2}}Pk_{-} & 0 & U+V & -\bar{S}_{-} & R & 0 &
\frac{1}{\sqrt{2}} \bar{S}_{-} & -\sqrt{2}R \\
 \sqrt{\frac{2}{3}}Pk_{z} & -\frac{1}{\sqrt{6}}Pk_{-} &
-\bar{S}^{\dag}_{-} & U-V & C & R & \sqrt{2}V &
-\sqrt{\frac{3}{2}}\tilde{S}_{-} \\
  \frac{1}{\sqrt{6}}Pk_{+} & \sqrt{\frac{2}{3}}Pk_{z} &
R^{\dag} & C^{\dag} & U-V & \bar{S}^{\dag}_{+} &
-\sqrt{\frac{3}{2}}\tilde{S}_{+} & -\sqrt{2}V \\
  0 & \frac{1}{\sqrt{2}}Pk_{+} & 0 & R^{\dag} & \bar{S}_{+} &
U+V & \sqrt{2}R^{\dag} & \frac{1}{\sqrt{2}}\bar{S}_{+} \\
  -\frac{1}{\sqrt{3}}Pk_{z} & -\frac{1}{\sqrt{3}}Pk_{-} &
\frac{1}{\sqrt{2}}\bar{S}^{\dag}_{-} & \sqrt{2}V &
-\sqrt{\frac{3}{2}}\tilde{S}^{\dag}_{+} & \sqrt{2}R & U-\Delta & C \\
  -\frac{1}{\sqrt{3}}Pk_{+} & \frac{1}{\sqrt{3}}Pk_{z} &
  -\sqrt{2}R^{\dag} & -\sqrt{\frac{3}{2}}\tilde{S}^{\dag}_{-} &
  -\sqrt{2}V & \frac{1}{\sqrt{2}}\bar{S}^{\dag}_{+} & C^{\dag} &
  U-\Delta
\end{pmatrix},
\end{equation}
where
\begin{eqnarray}\label{MatrEl}
  k_{\|}^{2} &=& k_{x}^{2}+k_{y}^{2},~~~~k_{\pm}=k_{x}\pm
  ik_{y},~~~~k_{z}=-i\partial/\partial z,\nonumber\\
  T &=& E_{c}(z)+\frac{\hbar^{2}}{2m_{0}}\left((2F+1)k_{\|}^{2}+k_{z}(2F+1)k_{z}\right),\nonumber\\
  U &=& E_{v}(z)-\frac{\hbar^{2}}{2m_{0}}\left(\gamma_{1}k_{\|}^{2}+k_{z}\gamma_{1}k_{z}\right),\nonumber\\
  V &=&         -\frac{\hbar^{2}}{2m_{0}}\left(\gamma_{2}k_{\|}^{2}-2k_{z}\gamma_{2}k_{z}\right),\nonumber\\
  R &=&         -\frac{\hbar^{2}}{2m_{0}}\left(\sqrt{3}\mu k_{+}^{2}-\sqrt{3}\bar{\gamma}k_{-}^{2}\right),\\
  \bar{S}_{\pm} &=&
  -\frac{\hbar^{2}}{2m_{0}}\sqrt{3}k_{\pm}\left(\{\gamma_{3},k_{z}\}+[\kappa,k_{z}]\right),\nonumber\\
  \tilde{S}_{\pm} &=&
  -\frac{\hbar^{2}}{2m_{0}}\sqrt{3}k_{\pm}\left(\{\gamma_{3},k_{z}\}-\frac{1}{3}[\kappa,k_{z}]\right),\nonumber\\
  C &=& \frac{\hbar^{2}}{m_{0}}k_{-}[\kappa,k_{z}]\nonumber.
\end{eqnarray}
$[A,B]=AB-BA$ is the usual commutator and $\{A,B\}=AB+BA$ is the
anti-commutator for the operators $A$ and $B$; $P$ is the Kane
momentum matrix element; $E_{c}(z)$ and $E_{v}(z)$ are the
conduction and valence band edges, respectively; $\Delta$ is the
spin-orbit splitting energy; and $\gamma_1$, $\gamma_2$,
$\gamma_3$, $\kappa$ and $F$ describe the coupling to the remote
bands and result in the $\mu$ and $\bar{\gamma}$ parameters
according to $\mu=1/2\left(\gamma_3-\gamma_2\right)$ and
$\bar{\gamma}=1/2\left(\gamma_3+\gamma_2\right)$. Only the terms
with non-spherical (cubic) symmetry in the Hamiltonian are
proportional to the warping parameter $\mu$. The case of $\mu=0$
corresponds to the axial approximation. The intrinsic inversion
asymmetry is neglected in the Hamiltonian because this effect is
very small in HgTe based structures.\cite{Weiler81} The band
structure parameters for HgTe and CdTe at $T = 0$~K are listed in
Table~\ref{Parametr}. The dependence of the band gap ($E_{g}$) of
\cmtx{} on the temperature and composition $x$ is determined from
the empirical expression according to Laurenti {\it et
al.}\cite{Laurenti90} The valence band offset between HgTe and
CdTe is taken to be equal to 570~meV at $T = 0$~K, in agreement
with recent experiments,\cite{Becker00} and is assumed to vary
linearly with $x$.\cite{Shih87}

\begin{table}[h]
\caption{\label{Parametr}Band structure parameters of HgTe and
CdTe at $T=0$~K. \cite{PhD,Zhang01}}
\begin{tabular}{|c|cccccccc|}
  \hline\hline
  ~ & ~~~~~$E_{g}$~~~~~ & ~~~~~$\triangle$~~~~~ &
~$E_{P}=\frac{2m_{0}P^{2}}{\hbar^{2}}$~ & ~~~$F$~~~ &
~~~$\gamma_1$~~~ & ~~~$\gamma_2$~~~ & ~~~$\gamma_3$~~~ & ~~~$\kappa$~ \\
 \hline
 HgTe & -0.303~eV & 1.08~eV & 18.8~eV & 0    & 4.1  &  0.5  & 1.3  & -0.4 \\
 CdTe & ~1.606~eV & 0.91~eV & 18.8~eV & -0.09& 1.47 & -0.28 & 0.03 & ~-1.31~ \\
 \hline\hline
\end{tabular}
\end{table}

So far, we have only considered the case of HgTe/\cmtx{} QW's with
(001) orientation. However, HgTe based structures have also been
investigated with orientations other than (001), for example,
(112) heterostructures.\cite{Becker00,Becker03} The electronic
properties of such systems depend strongly on the growth
direction. An extension of the model to QW's of a given $(kkl)$
orientation can be obtained using the approach of Los \textit{et
al.}\cite{Los96}: the set of basis functions
[Eqs.~(\ref{BasisSet})] is changed to a set which adopts the
symmetry of the problem, and thus the transformed Hamiltonian once
again has the form of Eq.~(\ref{Hamilt}), while the matrix
elements [Eqs.~(\ref{MatrEl})] contain additional terms depending
on the structure orientation.\cite{PhD} The exact formulas for
these terms are given in appendix A. Also strain and piezoelectric
effects can be included, as discussed in appendix B.

During the last two decades much attention has been paid to the
theoretical and experimental understanding of diluted magnetic
semiconductors (DMS), both in
bulk\cite{Furdyna88B,Kossut97,Averous91} as well as in
low-dimensional structures.\cite{Radantsev01,Averous91}
Extensively studied examples of this category are A$^{\rm
II}_{1-y}$Mn$_{y}$B$^{\rm VI}$ alloys, in which the group II
component is replaced randomly by the transition metal Mn.
\cite{Furdyna88} So far, most research on magnetic two-dimensional
structures has been done on wide gap DMS materials.
\cite{Scherbakov01,Camilleri01} In the present work we consider
QW's with magnetic ions (Mn) in narrow gap \mmty{}/\cmtx{} QW's.
The two-dimensional confinement in the DMS based layer combined
with the exchange interaction between localized Mn magnetic
moments and mobile band electrons make such structures quite
interesting candidates for the study of their electronic and
magnetic properties.

The band structure of \mmty{}/\cmtx{} QW's in the absence of a
magnetic field can be calculated similarly as that of non-magnetic
QW's, cf. the description above. The only difference is that the
band structure parameters for the well now depend on the Mn
concentration.\cite{Furdyna88}

\begin{figure} [h]
\includegraphics[width=12cm]{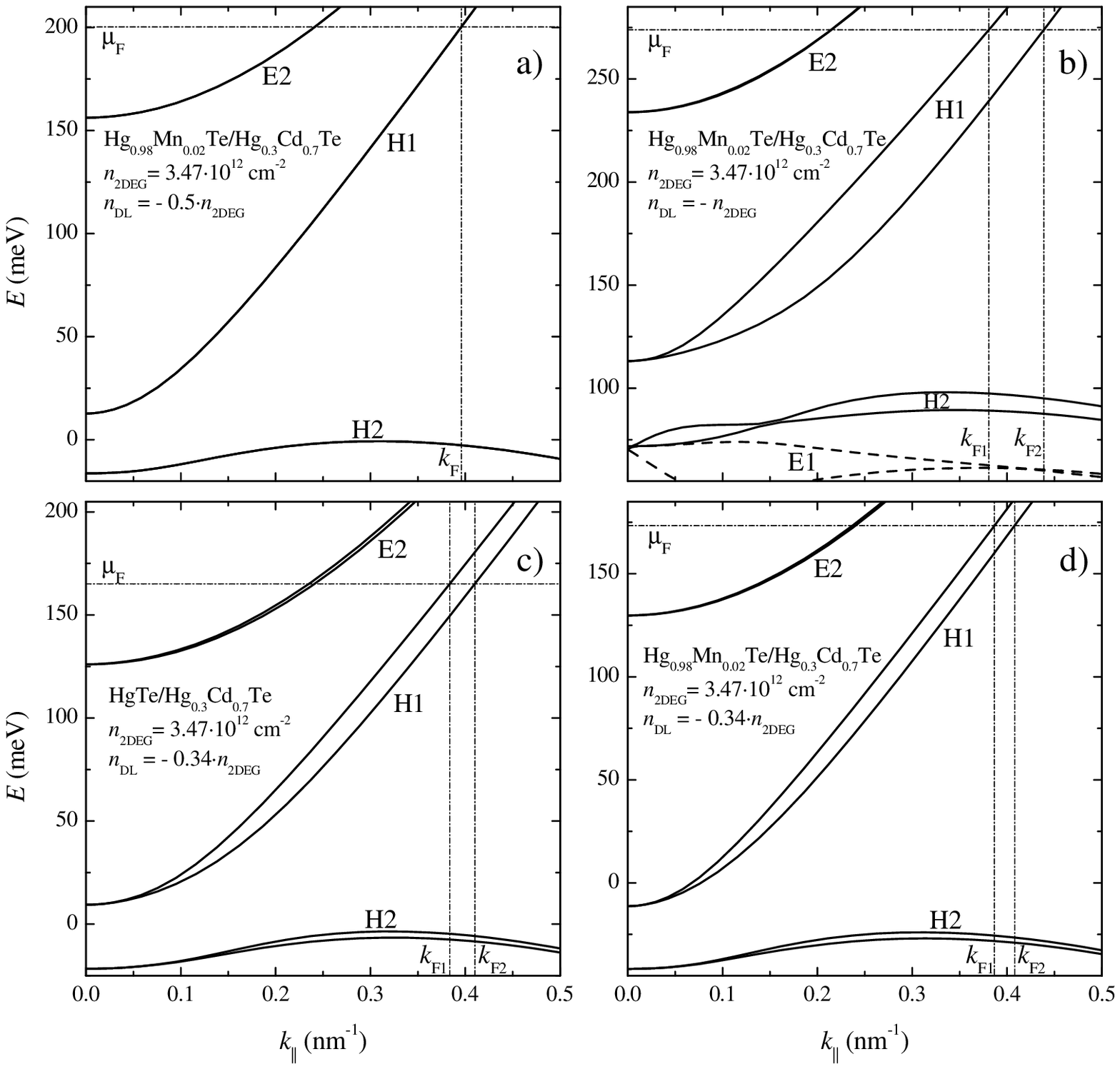}
\caption{\label{Rashba} Band structure of $n$-type (a) symmetrical
and (b,d) asymmetrical \mmt{}/\cmt{} QW's, as well as (c) an
asymmetrical HgTe/\cmt{} QW, all with $d_W = 12.2$~nm at $T =
4.2$~K. $k_{F1}=k_{F}(H1-)$ and $k_{F2}=k_{F}(H1+)$. Here the
$k_{\|}$ vector is $k_{\|}(1,0)$, however the difference between
$E(k_{\|}(1,0))$ and $E(k_{\|}(1,1))$ is less then 6~meV at
$k_{F}$ and the $k$-dependence shows qualitatively the same
behavior. The E1 subbands are shown as dashed lines.}
\end{figure}

In Fig.~\ref{Rashba} the zero-field subband dispersion $E(k_{\|})$
of $n$-type \mmt{}/\cmt{} (a, b, d) and HgTe/\cmt{} (c) (001) QW's
are presented. The temperature $T$ and the QW's width $d_W$ are
set at 4.2~K and 12.2~nm, respectively. Fig.~\ref{Rashba} (a)
corresponds to a \mmt{}/\cmt{} QW with symmetrically $n$-type
modulation doped barriers, while only the barrier on the substrate
side is doped for the case presented in Fig.~\ref{Rashba} (b). In
the calculations it is assumed that (i) all donors in the doped
layer are ionized, (ii) the charge density is constant in this
region, and (iii) all electrons are transferred to the QW. The
depletion charge in the doped layers is taken to be $n_{\rm
DL}=-n_{\rm 2DEG}$ for an asymmetrically and $n_{\rm
DL}=-1/2~n_{\rm 2DEG}$ for a symmetrically doped structure,
respectively.\cite{PhD} ($n_{\rm 2DEG}$ denotes the charge density
of the 2DEG in the QW and is chosen to be $n_{\rm 2DEG}=3.47\times
10^{12}$~cm$^{-2}$ for the calculations presented here.) The
eigenvalue [Eq.~(\ref{Eignprob})] and Poisson equations for the
two-dimensional charge carriers in the QW are solved
self-consistently for both cases. The depletion charge in the
doped layers, which is assumed to be fixed at the levels indicated
above, is included into the boundary conditions to solve the
Poisson equation.\cite{Ando82p} Then the Hartree potential is
determined according to the charge distribution of electrons in
the QW, which is given by the summation over all conduction band
states $i$ and all components $n$ of the envelope functions
$f_{n}(z)$:
\begin{equation}\label{HartreeE}
  \rho^{e}(z)=-e\sum_{i}^{\rm{CB}}\frac{1}{(2\pi)^2}\int\sum_{n=1}^{8}|f^{i}_{n}(z)|^2~f_{\rm{F}}(E_{i})~d^{2}k,
\end{equation}
where $e$ is the electron charge, $f_{\rm{F}}(E)$ is the Fermi
function. Here, we assume that $n_{\rm 2DEG}<0$ and $n_{\rm
DL}>0$. In analogy, we have for the holes:
\begin{equation}\label{HartreeE}
  \rho^{p}(z)=+e\sum_{i}^{\rm{VB}}\frac{1}{(2\pi)^2}\int\sum_{n=1}^{8}|f^{i}_{n}(z)|^2~(1-f_{\rm{F}}(E_{i}))~d^{2}k.
\end{equation}
In this case the summation index $i$ runs over all valence band
states, and $n_{\rm DL}<0$. The self-consistent Hartree potential
for zero magnetic field is then used to calculate the Landau
levels of the 2DEG (see description below).

From Fig.~\ref{Rashba} (a) and (b) one observes that both QW's
exhibit an inverted band structure. The conduction band includes
two occupied subbands (labeled as H1 and E2); the lower conduction
subband (H1) exhibits a heavy hole character at $k_{\|}=0$. The
electron-like E1 subband lies, in this case, below the H2 subband
and is now one of the valence subbands. The H1 and H2 subbands are
not split for the symmetric QW [case (a)]; however, for the
asymmetric QW, a spin-orbit splitting (in the following denoted as
a spin splitting as is common in the literature) of the H2 subband
as well as a pronounced splitting of the H1 subband are visible.
The splitting of H1 is 33.4~meV at $k_{F1}$ and 30.9~meV at
$k_{F2}$, respectively. The carrier concentrations in the H1$-$
and H1$+$ subbands for the asymmetric QW, after self-consistency
has been obtained, are $1.18\times 10^{12}$ and $1.55\times
10^{12}$~cm$^{-2}$, respectively. This large spin-orbit splitting,
usually called Rashba splitting, of the H1 state in an asymmetric
QW was demonstrated to be an unique feature of type-III QW's in
the inverted band regime.\cite{Zhang01} Experimentally values up
to 30~meV have indeed been observed.\cite{Gui-Giant}

The subband dispersion $E(k_{\|})$ of HgTe/\cmt{} and
\mmt{}/\cmt{} QW's are shown in Fig.~\ref{Rashba} (c) and (d),
where $n_{\rm DL}$ is taken to be equal to $-0.34\cdot n_{\rm
2DEG}$. The band structures of the non-magnetic (c) and magnetic
(d) QW's differ notably in the subband separation at $k_{\|}=0$;
the separation between the E2 and H1 subbands is about 24~meV
smaller for non-magnetic QW. The Rashba splitting of the H1
subband is 15.5~meV for the non-magnetic and 13.1~meV for the
magnetic QW's for $k_{\|}(1,0)$. The corresponding values are
14.2~meV and 12.2~meV for $k_{\|}(1,1)$. This relatively small
difference arises because the actual band gap
($\triangle_{\textrm{H1-H2}}$) changes only by $2\%$ upon
introduction of Mn.

\subsection{\label{sec:level2} $B\neq0$}

In an external magnetic field perpendicular to the plane of the
2DEG, the electronic bands are split into a series of Landau
levels. The effects of a magnetic field $\mathitb{B}=(0,0,B)$ can
be incorporated by a Peierls substitution
\cite{Pidgeon66,Luttinger55} in the Hamiltonian
[Eq.~(\ref{Hamilt})] as follows:
\begin{equation}\label{Peierls}
  \mathitb{k}\rightarrow\mathitb{k'}=-i\nabla+\frac{e}{\hbar}\mathitb{A},
\end{equation}
where $\mathitb{A}$ is the magnetic vector potential,
$\mathitb{B}=\nabla\times\mathitb{A}$. One possible Landau gauge
for $\mathitb{B}||z$ is $\mathitb{A}=(0,Bx,0)$. The operator
$\mathitb{k'}$ satisfies the following gauge-invariant relation:
\begin{equation}\label{kk}
  \mathitb{k'}\times\mathitb{k'}=-i\frac{e}{\hbar}\mathitb{B}.
\end{equation}
From now on, we drop the index $\mathitb{'}$ for simplicity.
Subsequently, $k_{x}$ and $k_{y}$ are rewritten such
that\cite{Pidgeon66,Weiler81}
\begin{equation}\label{CreaAnni}
  a= \frac{l_{c}}{\sqrt{2}}\cdot k_{-},~~~~
  a^{\dag}=\frac{l_{c}}{\sqrt{2}}\cdot k_{+},
\end{equation}
where $l_{c}=\sqrt{\frac{\hbar}{eB}}$ is the magnetic length; and
$a$ and $a^{\dag}$ are, respectively, the annihilation and
creation operators for the harmonic oscillator functions
$\varphi_{n}$, where\cite{Weiler81}
\begin{eqnarray}\label{oscill}
  a        \varphi_{n} &=& \sqrt{n}  \varphi_{n-1},\nonumber\\
  a^{\dag} \varphi_{n} &=& \sqrt{n+1}\varphi_{n+1}, \\
  a^{\dag}a\varphi_{n} &=&       n   \varphi_{n}.\nonumber
\end{eqnarray}
Here, $n=0,1,2,\ldots$ are the eigenvalues of the operator
$a^{\dag}a$. Thus, we can present the Hamiltonian in
Eq.~(\ref{Hamilt}) as a function of $a$, $a^{\dag}$,
$k_{z}=-i\partial/\partial z$, the band structure parameters and
their $z$-dependence.

Additionally, the Zeeman term,  $H^{Z}$, has to be included in the
Hamiltonian [Eq.~(\ref{Hamilt})]. As shown by
Weiler,\cite{Weiler81} this leads to the following matrix:
\begin{equation}\label{Zeeman}
H^{Z}=\hbar\frac{eB}{m_{0}}
\begin{pmatrix}
  \frac{1}{2} & 0 & 0 & 0 & 0 & 0 & 0 & 0 \\
  0 & -\frac{1}{2} & 0 & 0 & 0 & 0 & 0 & 0 \\
  0 & 0 & -\frac{3}{2}\kappa & 0 & 0 & 0 & 0 & 0 \\
  0 & 0 & 0 & -\frac{1}{2}\kappa & 0 & 0 & -\frac{\kappa+1}{\sqrt{2}} & 0 \\
  0 & 0 & 0 & 0 & \frac{1}{2}\kappa & 0 & 0 & -\frac{\kappa+1}{\sqrt{2}} \\
  0 & 0 & 0 & 0 & 0 & \frac{3}{2}\kappa & 0 & 0 \\
  0 & 0 & 0 & -\frac{\kappa+1}{\sqrt{2}} & 0 & 0 & -(\kappa+\frac{1}{2}) & 0 \\
  0 & 0 & 0 & 0 & -\frac{\kappa+1}{\sqrt{2}} & 0 & 0 & (\kappa+\frac{1}{2})
\end{pmatrix}.
\end{equation}

In the axial approximation we now assume that the total wave
function can be written as\cite{PhD}
\begin{equation}\label{Axial}
  \Psi_{N}(\mathitb{r})=\exp\left(-i\frac{X}{l_{c}^{2}}y\right)
\begin{pmatrix}
  & f_{1}(z) & \varphi_{n_1}~   \\
  & f_{2}(z) & \varphi_{n_2}~ \\
  & f_{3}(z) & \varphi_{n_3}~ \\
  & f_{4}(z) & \varphi_{n_4}~   \\
  & f_{5}(z) & \varphi_{n_5}~ \\
  & f_{6}(z) & \varphi_{n_6}~ \\
  & f_{7}(z) & \varphi_{n_7}~   \\
  & f_{8}(z) & \varphi_{n_8}~
\end{pmatrix}
  =\exp\left(-i\frac{X}{l_{c}^{2}}y\right)
\begin{pmatrix}
  & f_{1}^{(N)}(z) & \varphi_{N}~   \\
  & f_{2}^{(N)}(z) & \varphi_{N+1}~ \\
  & f_{3}^{(N)}(z) & \varphi_{N-1}~ \\
  & f_{4}^{(N)}(z) & \varphi_{N}~   \\
  & f_{5}^{(N)}(z) & \varphi_{N+1}~ \\
  & f_{6}^{(N)}(z) & \varphi_{N+2}~ \\
  & f_{7}^{(N)}(z) & \varphi_{N}~   \\
  & f_{8}^{(N)}(z) & \varphi_{N+1}~
\end{pmatrix},
\end{equation}
where restrictions on the quantum numbers $N$ on the
right-hand-side can be derived straightforwardly from
Eqs.~(\ref{oscill}). Since $n=0,1,2,\ldots$, the new quantum
number $N=-2,-1,0,\ldots$. For all quantum numbers $N$ a system of
(up to eight) coupled differential equations has to be solved. For
$N=-2$ the system is reduced to one equation that corresponds to a
state with purely heavy hole character. Non-axially symmetric
systems can be treated by taking the coupling between the
solutions of the axially symmetric problem [Eq.~(\ref{Axial})]
into account.\cite{Trebin79} The form of the coupling depends on
the symmetry along the growth direction and can be included by the
substitution of a linear combination of the $\Psi_{N}$ wave
functions:\cite{PhD}
\begin{equation}\label{Nonaxial}
  \Psi_{K}(\mathitb{r})=\sum_{N}c_{N}\Psi_{N}(\mathitb{r}),
\end{equation}
with $K=-2,-1,0,1$ and $N=K,K+4,K+8,\ldots$ for a (001) oriented
structure ($C_4$ symmetry); $K=-2,-1,0$ and $N=K,K+3,K+6,\ldots$
for a (111) structure ($C_3$ symmetry); $K=-2,-1$ and
$N=K,K+2,K+4,\ldots$ for a (110) structure ($C_2$ symmetry); and
$K=-2$ and $N=K,K+1,K+2,\ldots$ for other growth directions ($C_1$
symmetry). A system of coupled differential equations for the
envelope functions $f_{j}^{(N)}$ has to be solved for each value
of the quantum number $K$. The Hamiltonian matrix elements
$\langle\varphi_{n_{i}}|H_{ij}|\varphi_{n_{j}}\rangle$ are
determined using Eqs.~(\ref{CreaAnni}), (\ref{oscill}).

\begin{figure} [h]
\includegraphics[width=12cm]{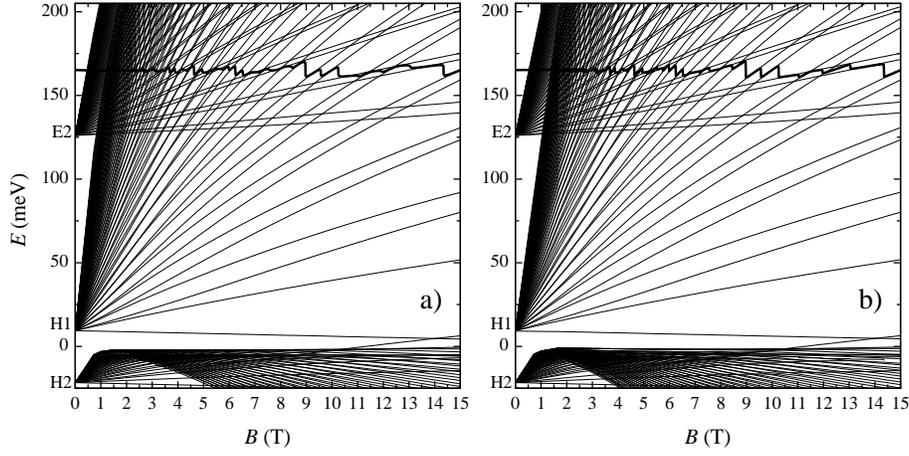}
\caption{\label{WithoutMn} Landau levels of the E2, H1 and H2
subbands for a $n$-type HgTe/\cmt{}(001) QW as a function of
magnetic field (a) with and (b) without applying the axial
approximation. $d_W = 12.2$~nm, $n_{\rm 2DEG} = 3.47\times
10^{12}$~cm$^{-2}$, $n_{\rm DL}=-0.34\cdot n_{\rm 2DEG}$, and $T =
4.2$~K. The thick line represents the chemical potential.}
\end{figure}

The Landau level spectra of a HgTe/\cmt{}(001) QW calculated with
and without applying the axial approximation are shown in
Fig.~\ref{WithoutMn} (a) and (b), respectively, for the structure
whose corresponding subband dispersion is presented in
Fig.~\ref{Rashba} (c). As far as the Landau level fan diagrams do
not show a notable difference we will use the axial approximation
in the following. As a result of the inverted band structure the
lowest Landau level of the H1 conduction subband and the highest
Landau level of the H2 valence subband cross at $B\approx14$~T.
Such a behavior is specific for type-III QW's and has been
examined theoretically and experimentally (see, for example,
Ref.~\onlinecite{Schultz98}). The lowest H1 Landau level, which
corresponds to the quantum number $N=-2$, has purely heavy hole
character, while the other Landau levels of the H1 subband are
mixed states. The H2 Landau level with $N=0$ contains both heavy
and light states. At $B>14$~T this level becomes the lowest level
of the conduction band.

\subsection{\label{sec:level2} Exchange interaction in magnetic {\mmty} QW's}

In the presence of a magnetic field, the $sp-d$ exchange
interaction of the $s$ and $p$ band electrons with the 3$d^{5}$
electrons of Mn in \mmty{} layer influences the band structure of
the QW. Such an interaction can be taken into account by adding an
appropriate exchange term ($H_{ex}$) to the Hamiltonian
[Eq.~(\ref{Hamilt})] in accordance with
Refs.~\onlinecite{Winkler03} and \onlinecite{Furdyna88}, which
leads to
\begin{equation}\label{Heisenberg}
  H+H_{ex}=H-\sum_{\mathitb{R}_{n}}J(\mathitb{r}-
\mathitb{R}_{n})\boldsymbol{\sigma}\mathitb{S}_{n},
\end{equation}
where $\boldsymbol{\sigma}$ is the spin operator of the band
electrons at the position $\mathitb{r}$, $\mathitb{S}_{n}$ is the
total spin operator of the $n$th Mn ion at position
$\mathitb{R}_{n}$, and $J(\mathitb{r}-\mathitb{R}_{n})$ is the
electron-ion exchange integral. Since the electron wave function
is extended, the spin operator $\mathitb{S}_{n}$ can be replaced
by the thermal average over all states of Mn moments $\langle
S_{z}\rangle$ for a magnetic field in the $z$-direction (mean
field approximation). Moreover, within the virtual crystal
approximation, $J(\mathitb{r}-\mathitb{R}_{n})$ can be replaced by
$yJ(\mathitb{r}-\mathitb{R})$, where $y$ is mole fraction of Mn,
and the summation is now carried out over all cation sites. The
exchange term in Eq.~(\ref{Heisenberg}) then
becomes\cite{Furdyna88}
\begin{equation}\label{VCA}
  H_{ex}=-\sigma_{z}\langle S_{z}\rangle y\sum_{\mathitb{R}}J
(\mathitb{r}-\mathitb{R}),
\end{equation}

The average $\langle S_{z}\rangle$ of the $z$ component of Mn spin
in the approximation of non-interacting magnetic moments is
determined by the empirical expression:\cite{Witowski82}
\begin{equation}\label{Mnmoment}
  \langle
  S_{z}\rangle=-S_{0}B_{5/2}\left(\frac{5~g_{Mn}\mu_{B}B}{2~k_{B}(T+T_{0})}\right),
\end{equation}
where $B_{5/2}(Z)$ is the Brillouin function for a spin of
$S=5/2$; $g_{Mn}=2$ is the $g$-factor of Mn; and the effective
spin $S_{0}$ and the effective temperature $(T+T_{0})$ account for
the existence of clusters and antiferromagnetic interaction
between Mn ions. The values for $S_{0}$ and $T_{0}$ are taken from
the literature.\cite{Gui04}

The matrix elements of $H_{ex}$ in terms of the Bloch functions
[Eqs.~(\ref{BasisSet})] have the form:
\begin{equation}\label{Hex}
H_{ex}=
\begin{pmatrix}
  3A\frac{\alpha}{\beta} & 0 & 0 & 0 & 0 & 0 & 0 & 0 \\
  0 & -3A\frac{\alpha}{\beta} & 0 & 0 & 0 & 0 & 0 & 0 \\
  0 & 0 & 3A & 0 & 0 & 0 & 0 & 0 \\
  0 & 0 & 0 & A & 0 & 0 & -2\sqrt{2}A & 0 \\
  0 & 0 & 0 & 0 & -A & 0 & 0 & -2\sqrt{2}A \\
  0 & 0 & 0 & 0 & 0 & -3A & 0 & 0 \\
  0 & 0 & 0 & -2\sqrt{2}A & 0 & 0 & -A & 0 \\
  0 & 0 & 0 & 0 & -2\sqrt{2}A & 0 & 0 & A
\end{pmatrix},
\end{equation}
with
\begin{equation}\label{AC}
  A=-\frac{1}{6}yN_{0}\beta\langle S_{z}\rangle.
\end{equation}
Here, $N_{0}$ is the number of unit cells per unit volume;
$\alpha$ and $\beta$ are constants which describe the exchange
interaction according to the $s-d$ and $p-d$ exchange integrals,
respectively. Experimental values for $\alpha$ and $\beta$ can be
found, for example, in Ref.~\onlinecite{Kossut97}.

The $sp-d$ exchange interaction changes the spin splitting of the
conduction and valence bands in a magnetic field. In the parabolic
approximation the effective $g$-factor for the $\Gamma_6$ states
can be described by the following equation [cf.
Eqs.~(\ref{Zeeman}), (\ref{Hex}) and (\ref{AC})]:
\begin{equation}\label{gfactorS}
  g_{eff}=g^{*}-\frac{yN_{0}\alpha\langle S_{z}\rangle}{\mu_{B}B},
\end{equation}
where $g^{*}$ is the $g$-factor of the band electrons (without
exchange term). The effect of the exchange interaction on the
$\Gamma_8$ states can be expressed by replacing the parameter
$\kappa$ with
\begin{equation}\label{gfactorP}
  \kappa_{eff}=\kappa+\frac{yN_{0}\beta\langle
  S_{z}\rangle}{6\mu_{B}B}.
\end{equation}

\begin{figure} [h]
\includegraphics[width=6.2cm]{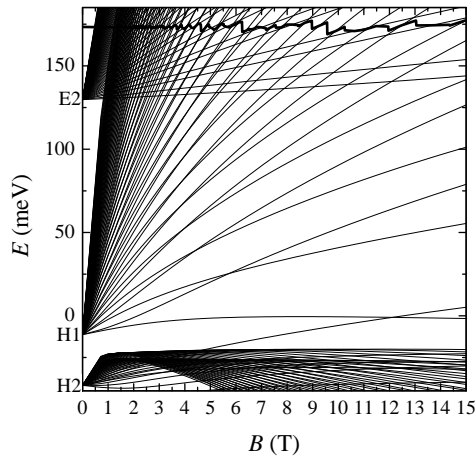}
\caption{\label{WithMn} Landau levels of the E2, H1 and H2
subbands for a $n$-type \mmt{}/\cmt{}(001) QW as a function of
magnetic field ($d_W = 12.2$~nm, $n_{\rm 2DEG}=3.47\times
10^{12}$~cm$^{-2}$, $n_{\rm DL}=-0.34\cdot n_{\rm 2DEG}$, and
$T$=4.2~K). The thick line represents the chemical potential.}
\end{figure}

The influence of the $sp-d$ exchange interaction on the band
structure is obvious when we compare the Landau levels in
Fig.~\ref{WithoutMn} for the non-magnetic structure with that in
Fig.~\ref{WithMn} for a magnetic \mmt{}/\cmt{}(001) QW. The QW
width, 2DEG density and temperature are the same in both cases.
The subband dispersion for the magnetic QW under consideration is
given in Fig.~\ref{Rashba} (d). The parameters $N_{0}\cdot \alpha
= 0.4$~eV, $N_{0}\cdot \beta = -0.6$~eV, $S_{0} = 5/2$ and $T_{0}
= 2.6$~K are taken from the literature.\cite{Gui04,Kossut97} Due
to the exchange interaction the lowest H1 Landau level with
quantum number $N=-2$ (which contains pure $|\Gamma_8,-3/2\rangle$
Bloch components) is bent upwards for low magnetic fields. In
contrast to the non-magnetic case, pairs of Landau levels from the
H1 subband cross even at moderate magnetic fields. At high
magnetic fields the ordering of the levels is the same as for
non-magnetic structures (Fig.~\ref{WithoutMn}). Such behavior was
also reported for $n$-type \mmty{} mixed
crystals.\cite{Jaczynski78} The crossing of the lowest Landau
level of the H1 subband with the $N=0$ Landau level of the H2
subband occurs at lower magnetic fields ($B\approx 12$~T) due to
the exchange enhanced shift towards higher energy of the H2 level.

In order to compare the calculations with experimental data, the
density of states (DOS) at the Fermi level has to be calculated
from the Landau level spectrum (Fig.~\ref{WithMn}), because
experimentally the Landau level structure becomes visible through
the magnetic field dependence of the longitudinal resistance. The
Shubnikov-de Haas (SdH) oscillations which are observed in the
experiments are directly related to changes of the DOS at the
Fermi energy. Assuming a Gaussian broadening of the Landau levels,
the DOS is given by\cite{Gerhardts76}
\begin{equation}\label{DOS}
 {\rm DOS}(E)=\frac{1}{2\pi l_c^2}\sum_{n}
 \frac{1}{\sqrt{\pi\Gamma^2}}\exp\left(-\frac{(E-E_{n})^2}{\Gamma^2}
 \right),
\end{equation}
where the summation runs over all Landau levels.
$\Gamma=\Gamma_{0}\sqrt{B/B_{0}}$ ($B_{0}$=1~T) is the Landau
level broadening parameter.\cite{Ando82}  In Fig.~\ref{DOS} the
calculated DOS for the Landau level spectrum of Fig.~\ref{WithMn}
is presented together with the SdH measurement of a
sample.\cite{Gui04} Growth and transport characterization
parameters (the QW width, the 2DEG density, the doping profile,
etc.) have been used for the band structure calculations. The
broadening parameter has been chosen to be $\Gamma_{0}$=1~meV. The
main features such as oscillation period, beating nodes and maxima
are in good agreement. For a more quantitative comparison the
magnetic field dependence of the diffusion constant had to be
taken into account.\cite{Ando82}

\begin{figure} [h]
\includegraphics[width=6.2cm]{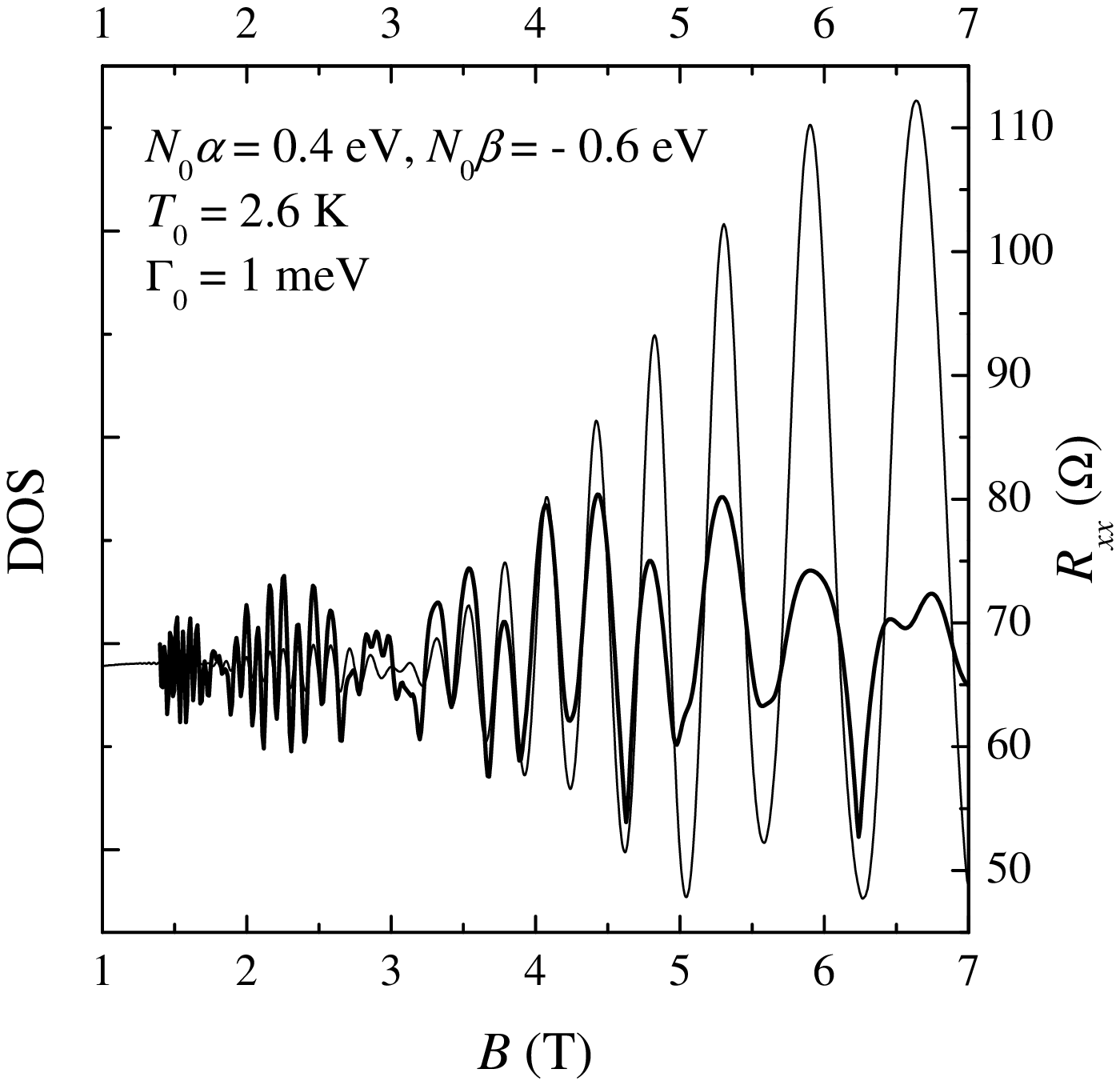}
\caption{\label{DOS} Density of states of the $n$-type
\mmt{}/\cmt{} QW at the Fermi level (thick line) compared with
experimental SdH oscillations (thin line).}
\end{figure}

\section{\label{sec:level1}Conclusion}

A detailed description has been presented of self-consistent band
structure calculations within an eight-band ${\mathitb k}\cdot
{\mathitb p}$ model in the envelope function approach with a
special emphasis on type-III HgTe/\cmtx{} QW structures. This
model is an important tool for the interpretation of both
optical\cite{Becker00,Becker03} and transport\cite{Zhang01}
experiments, where strong spin-orbit subband splitting effects are
observed. The model has been adopted to account for $sp - d$
exchange effects when magnetic (Mn) ions are introduced into the
structures. Self-consistently calculated band structure and DOS
are in good agreement with experimental transport results for
\mmt{}/\cmt{} QW's. The calculated band structure of
\mmty{}/\cmtx{} QW's and the ensuing comparison with experimental
data make it possible to understand the mutual influence of the
$sp-d$ exchange interaction and the two-dimensional confinement
effects on the transport properties. Moreover, the effect of the
QW parameters (width, doping profile, etc.) on the values of
$\alpha$, $\beta$, $S_{0}$ and $T_{0}$ can now be studied by a
direct comparison of experimental data and band structure
calculations.

\begin{acknowledgments}
We thank R.R.~Gerhardts, J.~Sinova and A.H.~MacDonald for fruitful
discussions. Financial support from the Deutsche
Forschungsgemeinschaft (SFB410 and GZ:436 WER) is gratefully
acknowledged.
\end{acknowledgments}

\appendix
\section{Corrections to the matrix elements of the Hamiltonian for
the [$kkl$] growth direction}

The approach of Los \textit{et al.} \cite{Los96} can be used to
carry out calculations for ($kkl$) oriented QW's. Since the
$\Gamma_6$ states as well as the coupling between $\Gamma_6$ and
$\Gamma_8$ ($\Gamma_7$) bands are spherically symmetric, only the
Bloch basis functions $u_{i}(\mathitb{r})$ ($i=3\ldots8$)
[Eqs.~(\ref{BasisSet})] have to be transformed into symmetry
adapted basis functions $u_{i}(\mathitb{r'})$. In addition, the
coordinate system is rotated to ($x'$,$y'$,$z'$) such that the
$z'$-axis is oriented along the [$kkl$] growth direction. The
corresponding terms are added to the matrix elements of
Eqs.~(\ref{MatrEl}). The corrections which depend on $k$ and $l$
($h=l/k$) are as follows (for simplicity the new coordinates are
referred as $x$, $y$, $z$):\cite{PhD}
\begin{eqnarray}\label{kklV}
  \Delta V     &=& \Delta V_{a}+\Delta V_{c},\nonumber\\
  \Delta V_{a} &=& -\frac{\hbar^{2}}{2m_{0}}\frac{6}{(h^{2}+2)^{2}}(2h^{2}+1)\left(\mu k_{\|}^{2}-2k_{z}\mu k_{z}\right),\\
  \Delta V_{c} &=& -\frac{\hbar^{2}}{2m_{0}}\frac{6}{(h^{2}+2)^{2}}(h^{2}-1)\left(\mu (k_{x}^{2}-k_{y}^{2})-h\sqrt{2}k_{x}\{\mu,k_{z}\}\right),\nonumber
\end{eqnarray}
\begin{eqnarray}\label{kklR}
  \Delta R     &=& \Delta R_{a}+\Delta R_{c},\nonumber\\
  \Delta R_{a} &=& \frac{\hbar^{2}}{2m_{0}}\frac{\sqrt{3}}{(h^{2}+2)^{2}}(2h^{2}+1)\mu k_{-}^{2},\\
  \Delta R_{c} &=& \frac{\hbar^{2}}{2m_{0}}\frac{\sqrt{3}}{(h^{2}+2)^{2}}\Bigl((2h^{4}+6h^{2}+1)\mu k_{+}^{2}+2(h^{2}-1)(\mu k_{\|}^{2}-2k_{z}\mu k_{z})\nonumber\\
               &+& h\sqrt{2}\left[(h^{2}+5)k_{+}-(h^{2}-1)k_{-}\right]\{\mu,k_{z}\}\Bigl),\nonumber
\end{eqnarray}
\begin{eqnarray}\label{kklS}
  \Delta S_{\pm}  &=& \Delta\bar{S}_{\pm}=\Delta\tilde{S}_{\pm}=\Delta S_{a\pm}+\Delta S_{c\pm},\nonumber\\
  \Delta S_{a\pm} &=& \frac{\hbar^{2}}{2m_{0}}\frac{4\sqrt{3}}{(h^{2}+2)^{2}}(2h^{2}+1)k_{\pm}\{\mu,k_{z}\},\\
  \Delta S_{c\pm} &=& \frac{\hbar^{2}}{2m_{0}}\frac{\sqrt{6}}{(h^{2}+2)^{2}}\Bigl(2h(h^{2}-1)(\mu k_{\|}^{2}-2k_{z}\mu k_{z})+h(h^{2}-1)\mu k_{\pm}^{2}\nonumber\\
                  &-& h(h^{2}+5)\mu k_{\mp}^{2}+2\sqrt{2}(h^{2}-1)k_{\mp}\{\mu,k_{z}\}\Bigl).\nonumber
\end{eqnarray}

The above terms are separated into axial (index $a$) and cubic
(index $c$) components. It can be shown that the axial and
non-axial approximations give the same result only for (001) and
(111) oriented structures at $k_{\|}=0$.

\section{Influence of strain and piezoelectric effects}

The effects of strain due to the lattice mismatch between HgTe and
\cmtx{} can be taken into consideration by applying a formalism
introduced by Bir and Pikus.\cite{Bir74} Terms proportional to the
strain tensor $\epsilon$ are added to the matrix elements of the
Hamiltonian [Eq.~(\ref{Hamilt})]; $H_{nn'}+H^{BP}_{nn'}$. The
Bir-Pikus Hamiltonian $H^{BP}$ is derived from Eq.~(\ref{Hamilt})
by the following substitution:
\begin{equation}\label{Tensor}
  k_{i}k_{j}\rightarrow \epsilon_{ij}.
\end{equation}
The strain tensor components ($\epsilon_{ij}$) transform as the
product $k_{i}k_{j}$ and are determined using the model of De~Caro
\textit{et al.}\cite{DeCaro95} The band structure parameters have
to be replaced by the deformation potentials;
\begin{eqnarray}\label{Potentials}
  \frac{\hbar^{2}}{2m_{0}}(2F+1) & \rightarrow & C,  \nonumber \\
  \frac{\hbar^{2}}{m_{0}}\gamma_{1} & \rightarrow & -2a, \nonumber\\
  \frac{\hbar^{2}}{m_{0}}\gamma_{2} & \rightarrow & -b, \\
  \frac{\hbar^{2}}{m_{0}}\gamma_{3} & \rightarrow & -\frac{1}{\sqrt{3}}d
  \nonumber.
\end{eqnarray}
Here, $C$ and $a$ are the hydrostatic, and $b$ and $d$ the
uniaxial deformation potentials. Due to the strain, the coupling
matrix elements between conduction ($\Gamma_6$) and valence
($\Gamma_8$, $\Gamma_7$) bands have additional terms which are
proportional to the Kane momentum matrix element
$P$.\cite{Trebin79} These elements are actually quite small and
consequently are neglected here. The Bir-Pikus Hamiltonian for
(001) oriented QW's can be written as\cite{PhD}
\begin{equation}\label{BirPikus}
H^{BP}=
\begin{pmatrix}
  T_{\epsilon} & 0 & 0 & 0 & 0 & 0 & 0 & 0 \\
  0 & T_{\epsilon} & 0 & 0 & 0 & 0 & 0 & 0 \\
  0 & 0 & U_{\epsilon}+V_{\epsilon} & S_{\epsilon} & R_{\epsilon} & 0 & -\frac{1}{\sqrt{2}}S_{\epsilon} & -\sqrt{2}R_{\epsilon} \\
  0 & 0 & S_{\epsilon}^{\dag} & U_{\epsilon}-V_{\epsilon} & 0 &  R_{\epsilon} & \sqrt{2}V_{\epsilon} & \sqrt{\frac{3}{2}}S_{\epsilon} \\
  0 & 0 & R_{\epsilon}^{\dag} & 0 & U_{\epsilon}-V_{\epsilon} & -S_{\epsilon} & \sqrt{\frac{3}{2}}S_{\epsilon}^{\dag} & -\sqrt{2}V_{\epsilon} \\
  0 & 0 & 0 & R_{\epsilon}^{\dag} & -S_{\epsilon}^{\dag} & U_{\epsilon}+V_{\epsilon} & \sqrt{2}R_{\epsilon}^{\dag} & -\frac{1}{\sqrt{2}}S_{\epsilon}^{\dag} \\
  0 & 0 & -\frac{1}{\sqrt{2}}S_{\epsilon}^{\dag} & \sqrt{2}V_{\epsilon} & \sqrt{\frac{3}{2}}S_{\epsilon} & \sqrt{2}R_{\epsilon} & U_{\epsilon} & 0 \\
  0 & 0 & -\sqrt{2}R_{\epsilon}^{\dag} & \sqrt{\frac{3}{2}}S_{\epsilon}^{\dag} & -\sqrt{2}V_{\epsilon} & -\frac{1}{\sqrt{2}}S_{\epsilon} & 0 & U_{\epsilon}
\end{pmatrix},
\end{equation}
where
\begin{eqnarray}
  T_{\epsilon} & = & C~tr(\epsilon), \nonumber \\
  U_{\epsilon} & = & a~tr(\epsilon), \nonumber \\
  V_{\epsilon} & = & \frac{1}{2}b(\epsilon_{xx}+\epsilon_{yy}-2\epsilon_{zz}), \\
  S_{\epsilon} & = & -d(\epsilon_{xz}-i\epsilon_{yz}), \nonumber \\
  R_{\epsilon} & = & -\frac{\sqrt{3}}{2}b(\epsilon_{xx}-\epsilon_{yy})+id\epsilon_{xy}. \nonumber
\end{eqnarray}
$tr(\epsilon)=\epsilon_{xx}+\epsilon_{yy}+\epsilon_{zz}$ is the
trace of the strain tensor.

For ($kkl$) oriented structures the Hamiltonian should be
presented in the symmetry adapted set of basis functions as
described in appendix A. The transformed Hamiltonian has the form
of Eq.~(\ref{BirPikus}), with appropriate corrections to the
matrix elements. These corrections can be derived from
Eqs.~(\ref{kklV}), (\ref{kklR}) and (\ref{kklS}) by the
substitutions indicated in Eqs.~(\ref{Tensor}) and
(\ref{Potentials}).

If the strain tensor has non-zero off-diagonal components (shear
components), internal electric fields are generated in the QW due
to the piezoelectric effect. We have calculated the strain-induced
polarization and electric fields as described in
Ref.~\onlinecite{DeCaro95}, and have found that the influence of
piezoelectric fields on the band structure of fully strained
HgTe/\cmtx{} (112) heterostructures is negligible.\cite{Becker00}




\end{document}